\documentclass[epj]{webofc}
\usepackage[varg]{txfonts}   
\usepackage{graphicx,color} 
\usepackage{bm}       
\usepackage{amsmath}
\usepackage{amssymb}
\def\Journal#1#2#3#4{{#1} {\bf#2}, #3 (#4)}
\def\Journal#1#2#3#4{{#1} {\bf#2}, #3 (#4)}

\def\PLB{{\rm Phys. Lett.}  B}
\def\PRL{\rm Phys. Rev. Lett.}
\def\PRD{{\rm Phys. Rev.} D}

\def\la{\langle}
\def\ra{\rangle}

\def\al{\alpha}

\def\be{\begin{equation}}
\def\ee{\end{equation}}
\def\bea{\begin{eqnarray}}
\def\eea{\end{eqnarray}}
\woctitle{21st International Conference on Few-Body Problems in Physics}
\begin{document}
\title{Light-Front Quark Model Phenomenology Consistent with Chiral Symmetry}
\author{Ho-Meoyng Choi\inst{1}\fnsep\thanks{\email{homyoung@knu.ac.kr}} \and
        Chueng-Ryong Ji\inst{2}\fnsep\thanks{\email{crji@ncsu.edu}}
}

\institute{
Department of Physics, Teachers College, Kyungpook National University, Daegu 702-701, Korea
\and
Department of Physics, North Carolina State University, Raleigh, NC 27695-8202, USA}

\abstract{ We discuss the link between the chiral symmetry of QCD and the numerical
results of the light-front quark model, analyzing both the
twist-2 and twist-3 distribution amplitudes of a pion.
In particular, we exhibit that the pion twist-3 distribution amplitude can flip the sign of curvature
as the cutoff scale for the transverse momentum gets lower than a typical scale of around 1 GeV.
}
\maketitle
\section{Introduction}
\label{intro}
Hadronic distribution amplitudes (DAs) provide essential
information on the QCD interaction of quarks, antiquarks and gluons inside the hadrons and
play an essential role in applying QCD to hard exclusive processes~\cite{LB}.
It has motivated many studies using various nonperturbative models
and led to develop distinct phenomenological models over the past two decades.
Among them, the light-front quark model (LFQM) appears to be one of the most efficient and effective tools in hadron physics as it takes advantage of the distinguished features of the LFD.
The rational energy-momentum dispersion relation of LFD yields the sign correlation between the LF energy $k^-(=k^0-k^3)$ and
the LF longitudinal momentum $k^+(= k^0 + k^3)$ and leads to the suppression of vacuum fluctuations.
This simplification is a remarkable advantage in LFD and facilitates the partonic interpretation of the amplitudes. Based on the advantages of the LFD, the LFQM  has been developed and subsequently applied for various meson phenomenologies such as the meson mass spectra, the decay constants, DAs, form factors and GPDs~\cite{CJ_99,MF12}. Despite these successes in reproducing the general features
of the data, however, it has proved very difficult to obtain direct connection between the LFQM and QCD.

Through the recent analysis of the twist-2 and twist-3 DAs of pseudoscalar and vector
mesons~\cite{CJ14,CJ15}, we discussed the link between the chiral symmetry of QCD and the numerical
results of the LFQM. In particular, through the analysis of twist-3 DAs of $\pi$ and $\rho$ mesons, we observe that the LFQM with effective degrees of freedom represented by the constituent quark and antiquark may provide the view of effective zero-mode cloud around the quark and antiquark inside the meson. Consequently, the constituents dressed by the zero-mode cloud may be expected to satisfy the chiral symmetry of QCD. Our numerical results appear consistent with this expectation and effectively indicate that the constituent quark and antiquark in the LFQM may be considered as the dressed constituents including the zero-mode quantum fluctuations from the vacuum.


\section{Model Description}
\label{sec-1}
The twist-2(3) DAs $\phi^{\cal A(P)}_{2(3);\pi}$  for a pion
are defined in the light-cone gauge as follows~\cite{DA1}:
\bea\label{Deq:1}
\la 0|{\bar q}(z)\gamma^\mu\gamma_5 q(-z)|\pi(P)\ra
&=& if_\pi P^\mu \int^1_0 dx e^{i(2x-1) P\cdot z} \phi^{\cal A}_{2;\pi}(x),
\nonumber\\
\la 0|{\bar q}(z)i\gamma_5 q(-z)|\pi(P)\ra
&=& {f_\pi}\mu_\pi \int^1_0 dx e^{i(2x-1) P\cdot z} \phi^{\cal P}_{3;\pi}(x),
\eea
where
the normalization parameter $\mu_\pi$
is given by $\mu_\pi = -2\la {\bar q}q\ra / f^2_\pi$ from the Gell-Mann-Oakes-Renner relation.
Defining the local matrix elements
${\cal M}_\al \equiv \la 0|{\bar q}\Gamma_\al q|\pi(P)\ra~ (\al={\cal A, P})$
for axial-vector ($\Gamma_A =\gamma^\mu\gamma_5$) and pseudoscalar ($\Gamma_P =i\gamma_5$) channels, we start from the exactly solvable manifestly covariant Bethe-Salpter (BS) model:
%
\be\label{Deq:4}
{\cal M}_\al = N_c
\int\frac{d^4k}{(2\pi)^4} \frac{H_0} {(p^2 -m^2_q +i\varepsilon)(k^2 - m^2_{\bar q}+i\varepsilon)}
{\rm Tr}\left[\Gamma_\al\left(\slash \!\!\!p+m_q \right)
\gamma_5
\left(-\slash \!\!\!k + m_{\bar q} \right) \right],
\ee
where $N_c$ denotes the number of colors, $m_{q({\bar q})}$ is the constituent quark~(antiquark) mass, and $p =P -k$.
In order to regularize the covariant loop,
we use the usual multipole ansatz~$H_0(p^2,k^2) = g / (p^2 - \Lambda^2 +i\varepsilon)^2$
for the $q{\bar q}$ bound-state vertex function,
where $g$ and $\Lambda$ are constant parameters.
Since the vertex function $H_0$
may not be the realistic approximation of a $q\bar{q}$ bound state, we
utilize it only as a tool to analyze the zero-mode complication in the exactly solvable BS model.
Comparing the two results of Eq.~(\ref{Deq:4}) obtained from the manifestly covariant calculation
and the LF one, we found that only the LF result for ${\cal M}_{\cal A}$ receives the zero modes.
What is remarkable in our finding~\cite{CJ14,CJ15} is that the nonvanishing
zero-mode contributions as well as the instantaneous
ones to ${\cal M}_{\cal A}$ that appeared in the covariant BS model now vanish explicitly
when the phenomenological wave function such as the
Gaussian wave function in the LFQM is used. In other words,
the decay constants and the quark DAs of pseudoscalar mesons
can be obtained only from the on-mass-shell valence
contribution within the framework of the standard LFQM  using the Gaussian radial wave
function and they still satisfy the chiral symmetry consistent with the QCD.
This observation is also applicable to the calculation of the pion elastic form factor~\cite{CJ15}.
The self-consistent correspondence relation
between the covariant BS model and the LFQM can be found in~\cite{CJ14,CJ15}.

The explicit forms of the twist-2 and twist-3 DAs of the pion in the LFQM are given by~\cite{CJ15}
\bea\label{QM9}
\phi^{\cal A}_{2;\pi}(x) &=& \frac{\sqrt{2N_c}}{{f_{\pi}8\pi^3}}
\int^{|{\bf k}_\perp|<\mu} d^2{\bf k}_\perp
\frac{\phi_R(x,{\bf k}_\perp)}{\sqrt{{\bf k}^2_\perp + m^2_q}} m_q
\equiv \int^{|{\bf k}_\perp|<\mu} d^2{\bf k}_\perp
\psi^{\cal A}_{2;\pi}(x, {\bf k}_\perp),
\nonumber\\
 \phi^{\cal P}_{3;\pi}(x)
 &=& -\frac{\sqrt{2N_c}}{f_\pi\mu_\pi \cdot 16 \pi^3}
 \int^{|{\bf k}_\perp|<\mu} d^2{\bf k}_\perp
 \frac{\phi_R(x,{\bf k}_\perp)}{\sqrt{{\bf k}^2_\perp + m^2_q}} M^2_0
 \equiv \int^{|{\bf k}_\perp|<\mu} d^2{\bf k}_\perp
\psi^{\cal P}_{3;\pi}(x, {\bf k}_\perp),
\eea
respectively. We can also define the ${\bf k}_\perp^2$ distribution
$\psi^{\cal A(P)}_{2(3);\pi}({\bf k}_\perp^2)$ after integrating first over $x$ for the full
wave function $\psi^{\cal A(P)}_{2(3);\pi}(x, {\bf k}_\perp)$.
From the point of view of QCD, {\color{black} one should note that} the quark DAs of a hadron depend on the scale $\mu$ that {\color{black} may separate} nonperturbative and perturbative regimes. In our LFQM, we can associate $\mu$
with the transverse integration cutoff via $|{\bf k}_\perp|\leq \mu$.
The dependence on the scale $\mu$ is then {\color{black} consistently} given by the QCD evolution equation~\cite{LB}.
In our previous analysis~\cite{CJ15} for the radial wave function $\phi_R(x, {\bf k}_\perp)$,
we used the Gaussian or Harmonic Oscillator~(HO) wave function,
$\phi^{\rm HO}_R=N_{\rm HO} \exp(-\vec{k}^2/2\beta^2)$
where $\vec{k}^2={\bf k}^2_\perp + k^2_z = {\bf k}^2_\perp +(x-1/2)^2 M_0^2$ for the pion case.
But, in this work, we also use the power-law (PL) wave function,
$\phi^{\rm PL}_R= N_{\rm PL} (1 + {\vec k}^2/\beta^2)^{-2}$ and
check the sensitivity of the scale $\mu$ to $\phi^{\cal A}_{2;\pi}(x)$
and $\phi^{\cal P}_{3;\pi}(x)$.

\section{Numerical Results}
In our numerical calculations, we use the model parameters
$m_{u(d)}=0.22$ GeV and $\beta = 0.3659$ GeV for both $\phi^{\rm HO}_R$ and $\phi^{\rm PL}_R$.
Both $\phi^{\rm HO}_R$ and $\phi^{\rm PL}_R$ are normalized without
the momentum cutoff (i.e. $|{\bf k}_\perp|\to\infty$) and the corresponding pion
decay constants give the same value $f_\pi=130$ MeV, which is
in good agreement with the experimental data~, $f^{\rm Exp.}_\pi=(130.41\pm 0.03 \pm 0.20)$ MeV.

\begin{figure}
\centering
\sidecaption
\includegraphics[width=5cm,clip]{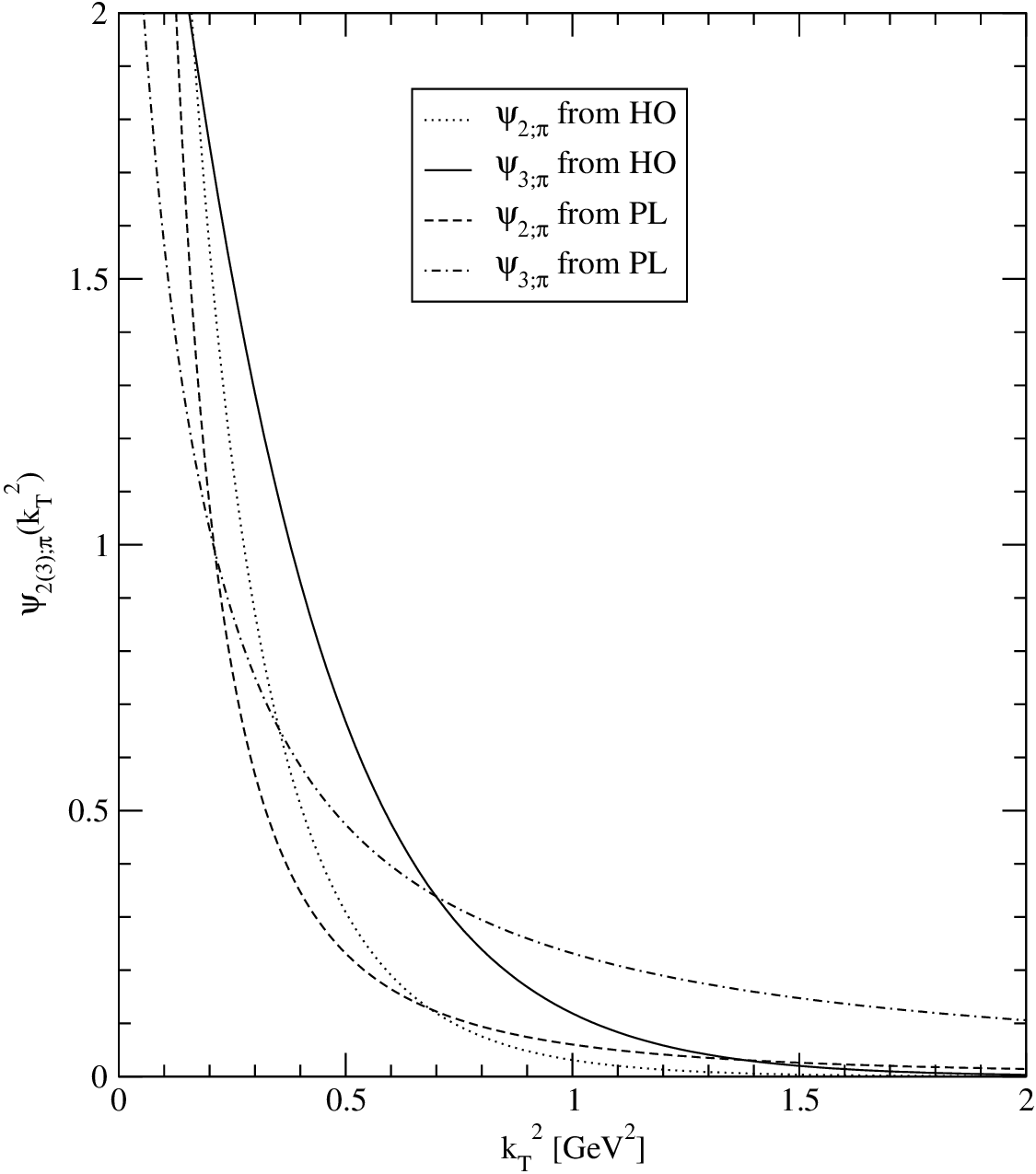}
\caption{
The ${\bf k}_\perp^2$ distributions
$\psi^{\cal A(P)}_{2(3);\pi}({\bf k}_\perp^2)$ obtained from
HO and PL wave functions.
The dotted, solid, dashed, dot-dashed lines represent
$[\psi^{\cal A}_{2;\pi}]_{\rm HO}$, $[\psi^{\cal P}_{3;\pi}]_{\rm HO}$,
$[\psi^{\cal A}_{2;\pi}]_{\rm PL}$, and
$[\psi^{\cal P}_{3;\pi}]_{\rm PL}$, respectively.
}
\label{fig-1}       
\end{figure}
\begin{figure}
\centering
\includegraphics[width=5cm,clip]{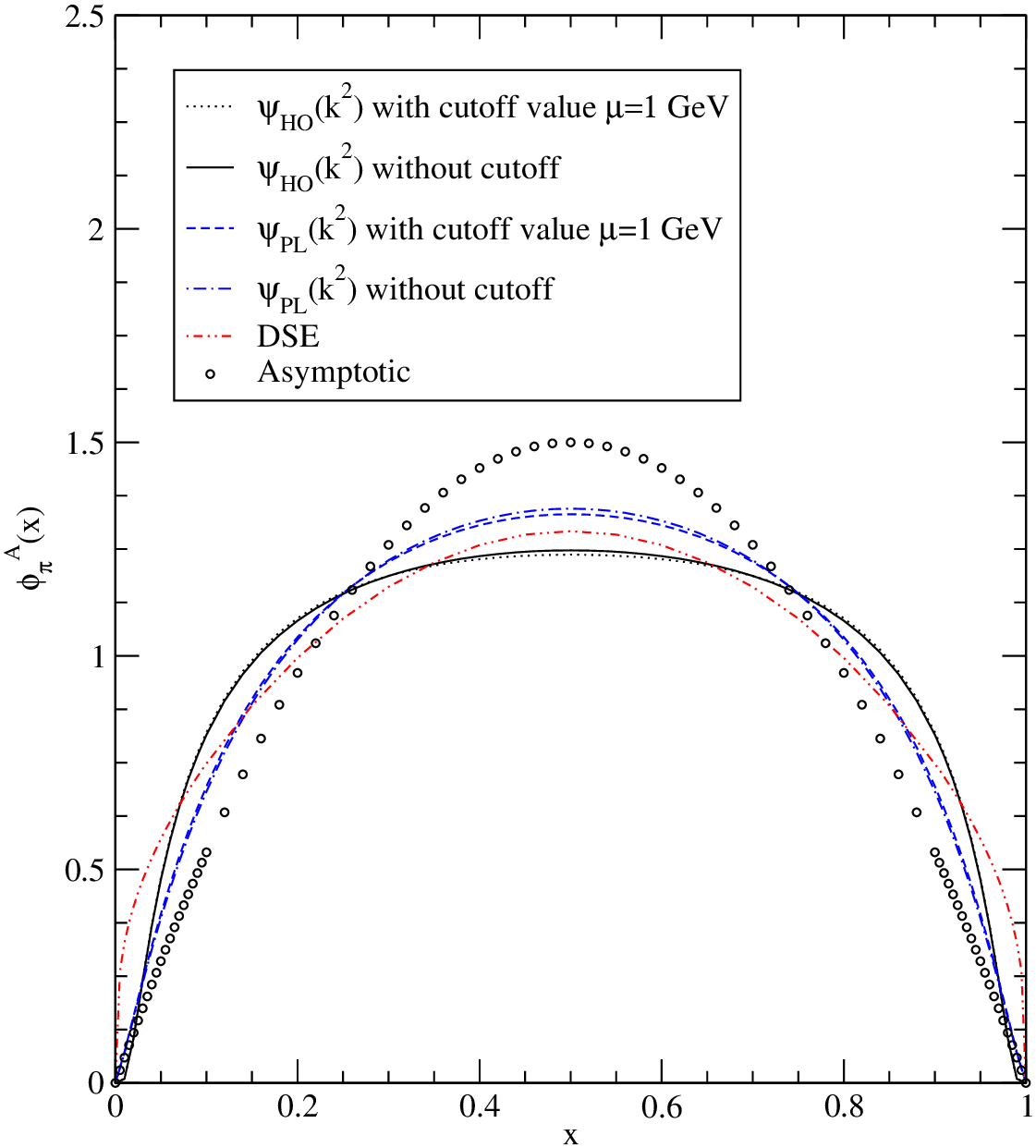}
\hspace{1cm}
\includegraphics[width=5cm,clip]{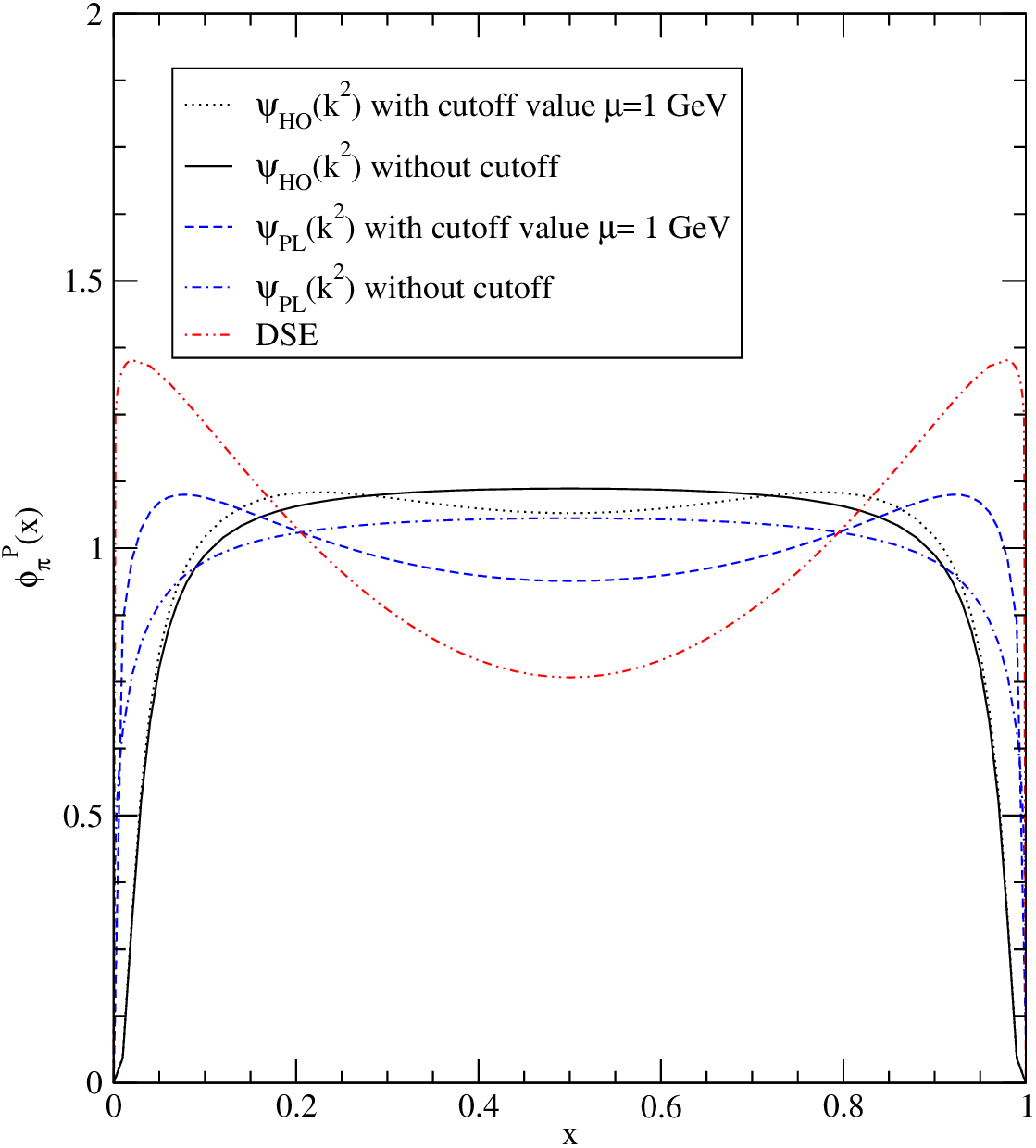}
\caption{Twist-2 and-3 DAs from Gaussian (left) vs. Power-law (right) wave function.}
\label{fig-2}       
\end{figure}

Figure~\ref{fig-1} shows the ${\bf k}_\perp^2$ distributions
$\psi^{\cal A(P)}_{2(3);\pi}({\bf k}_\perp^2)$ obtained from
both $\phi^{\rm HO}_R$ and $\phi^{\rm PL}_R$ .
As one can see from Fig.~\ref{fig-1},
$[\psi^{\cal A(P)}_{2(3);\pi}]_{\rm PL}$ have more high momentum tails
than the corresponding $[\psi^{\cal A(P)}_{2(3);\pi}]_{\rm HO}$
for the cutoff values $\mu\geq 1$ GeV.
Figure~\ref{fig-2} shows $\phi^{\cal A}_{2;\pi}(x)$ (left panel)
and $\phi^{\cal P}_{3;\pi}(x)$ (right panel) of a pion. We plotted for the two cases, i.e.
without the cutoff (i.e. $\mu\to\infty$) and with the cutoff value $\mu\to 1$ GeV for
both $\phi^{\rm HO}_R$ and $\phi^{\rm PL}_R$. As one see from Fig.~\ref{fig-2},
$\psi^{\cal P}_{3;\pi}(x)$ for both $\phi^{\rm HO}_R$ and $\phi^{\rm PL}_R$ are more sensitive to the
cutoff scale than $\psi^{\cal A}_{2;\pi}(x)$.
While $\psi^{\cal P}_{3;\pi}$ for both HO and PL wave functions
show slight convex shapes when $\mu\to\infty$, they flip the convex shapes to concave ones when
the typical momentum cutoff scale $\mu=1$ GeV or less is taken. The concave-shape $\psi^{\cal P}_{3;\pi}(x)$ was obtained in other models
such as the QCD sum rules~\cite{DA1} and Dyson-Schwinger approach~\cite{DA2} (double-dot dashed lines)
and our LFQM calculation can generate the concave-shape with a finite transverse momentum cutoff.
This concave behaviors of $\psi^{\cal P}_{3;\pi}(x)$ are more pronounced
for the smaller cutoff scale as one may expect from Fig.~\ref{fig-1}.
This rebuts the remark made in Ref.\cite{Craig2015} that our LFQM has curvature of the opposite sign
on almost the entire domain of support in conflict with a model-independent prediction of QCD.
We have shown in our recent works\cite{CJ14,CJ15} that our LFQM is indeed consistent with the nature of chiral symmetry in QCD.
 Without the momentum cutoff ($\mu\to\infty$), our results for
both $\phi^{\rm HO}_R$ and $\phi^{\rm PL}_R$
reproduce the asymptotic result~\cite{BF},
$[\phi^{\cal P}_{3;\pi}]_{\rm as}(x)\to 1$,
in the chiral symmetry ($m_q\to 0$) limit in our LFQM.
Our LFQM result
in the chiral symmetry limit is consistent with the conclusion drawn from our previous
analysis~\cite{CJ14} of the twist-2 ($\phi^{||}_{2;\rho}(x)$)
and twist-3 ($\phi^{\perp}_{3;\rho}(x)$) $\rho$ meson DAs, where both DAs also
remarkably reproduce the exact asymptotic DAs in the chiral symmetry limit. However, with the
momentum cutoff, the asymptotic behavior of $\phi^{\cal P}_{3;\pi}(x)$ becomes concave shape
following the results of  the QCD sum rules~\cite{DA1} and Dyson-Schwinger approach~\cite{DA2}.

\section{Summary}
We have discussed the link between the chiral symmetry of QCD and the numerical
results of the LFQM, analyzing both the twist-2 and twist-3 DAs of a pion using
two different model wave functions~(HO vs. PL).
While the twist-2 DA is rather insensitive to the transverse momentum cutoff $\mu\geq 1$ GeV,
the twist-3 DA is sensitive to the cutoff value and thus the sign of curvature should be
discussed discretely according to the scale taken consistently with the nature of chiral symmetry in QCD.
We also note that the result $[\phi^{\cal P}_{3;\pi}]_{\rm as}(x)\to 1$ for the
twist-3 DA in the chiral symmetry limit is independent of the model wave functions when the
$\mu\to\infty$.

\begin{acknowledgement}
This work is supported by the Korean Research Foundation~(No. NRF-2014R1A1A2057457).
CRJ acknowledges partial support from the US Department of Energy~
(No. DE-FG02-03ER41260).
\end{acknowledgement}

%
%
%

\end{document}